\documentclass[preprint]{aastex}
\usepackage{epsf}
\usepackage{emulateapj5}
\usepackage{onecolfloat}
\usepackage{apjfonts}
\usepackage{amsmath}

\def \xoff {\ifmmode x_{\rm off} \else $x_{\rm off}$ \fi}
\def \rhorms {\ifmmode \rho_{\rm rms} \else $\rho_{\rm rms}$ \fi}

\def \apj  {ApJ}
\def \apjs  {ApJS}

\def \mnras {MNRAS}
\def \etal {et~al.~}
\def \chisq  {\ifmmode  \chi^2   \else  $\chi^2$  \fi}  
\def \chisqr {\ifmmode \chi^2_{\rm r} \else $\chi^2_{\rm r}$ \fi}
\def \spose#1{\hbox  to 0pt{#1\hss}}  
\def \lta{\mathrel{\spose{\lower 3pt\hbox{$\sim$}}\raise  2.0pt\hbox{$<$}}}
\def \gta{\mathrel{\spose{\lower  3pt\hbox{$\sim$}}\raise 2.0pt\hbox{$>$}}}
 
\def \ha  {\ifmmode H\alpha \else H$\alpha $ \fi}

\def \kms {\ifmmode  \,\rm km\,s^{-1} \else $\,\rm km\,s^{-1}  $ \fi }
\def \kpc {\ifmmode  {\rm kpc}  \else ${\rm  kpc}$ \fi  }  
\def \Msun {\ifmmode M_{\odot} \else $M_{\odot}$ \fi} 
\def \hMsun {\ifmmode h^{-1}\,\rm M_{\odot} \else $h^{-1}\,\rm M_{\odot}$ \fi}
\def \hhMsun {\ifmmode h^{-2}\,\rm M_{\odot}\else $h^{-2}\,\rm M_{\odot}$ \fi}
\def \Lsun {\ifmmode L_{\odot} \else $L_{\odot}$ \fi} 
\def \hhLsun {\ifmmode h^{-2}\,\rm L_{\odot} \else $h^{-2}\,\rm L_{\odot}$ \fi}

\def \LCDM {\ifmmode \Lambda{\rm CDM} \else $\Lambda{\rm CDM}$ \fi}
\def \sig8 {\ifmmode \sigma_8 \else $\sigma_8$ \fi} 
\def \OmegaM {\ifmmode \Omega_{\rm M} \else $\Omega_{\rm M}$ \fi} 
\def \OmegaL {\ifmmode \Omega_{\rm \Lambda} \else $\Omega_{\rm \Lambda}$\fi} 
\def \Deltavir {\ifmmode \Delta_{\rm vir} \else $\Delta_{\rm vir}$ \fi}

\def \rs {\ifmmode r_{\rm s} \else $r_{\rm s}$ \fi} 
\def \rrm2 {\ifmmode r_{-2} \else $r_{-2}$ \fi} 
\def \ccm2 {\ifmmode c_{-2} \else$c_{-2}$ \fi} 
\def \cvir {\ifmmode c_{\rm vir} \else $c_{\rm vir}$ \fi} 
\def \cbar {\ifmmode \overline{c} \else $\overline{c}$ \fi}

\def \R200 {\ifmmode R_{200} \else $R_{200}$ \fi} 
\def \Rvir {\ifmmode R_{\rm vir} \else $R_{\rm vir}$ \fi}

\def \v200 {\ifmmode V_{200} \else $V_{200}$ \fi} 
\def \Vvir {\ifmmode V_{\rm  vir} \else  $V_{\rm vir}$  \fi} 
\def  \Vhalo  {\ifmmode V_{\rm halo} \else $V_{\rm halo}$ \fi}

\def \M200 {\ifmmode M_{200} \else $M_{200}$ \fi} 
\def \Mvir {\ifmmode M_{\rm  vir} \else $M_{\rm  vir}$ \fi}  
\def \Mshell  {\ifmmode M_{\rm shell} \else $M_{\rm shell}$ \fi}

\def \Nvir {\ifmmode N_{\rm  vir} \else $N_{\rm  vir}$ \fi}  

\def \Jvir {\ifmmode J_{\rm vir} \else $J_{\rm vir}$ \fi} 
\def \Jshell {\ifmmode J_{\rm shell} \else $J_{\rm shell}$ \fi}

\def \Evir {\ifmmode E_{\rm vir} \else $E_{\rm vir}$ \fi} 

\def \lam {\ifmmode \lambda  \else $\lambda$ \fi} 
\def \lamp {\ifmmode \lambda^{\prime} \else $\lambda^{\prime}$  \fi} 
\def \lampc {\ifmmode \lambda^{\prime}_{\rm c} \else
  $\lambda^{\prime}_{\rm c}$  \fi} 
\def \lambar {\ifmmode \bar{\lambda}  \else  $\bar{\lambda}$  \fi}  
\def  \lampbar  {\ifmmode \bar{\lambda^{\prime}} \else
  $\bar{\lambda^{\prime}}$\fi} 
\def \siglam {\ifmmode \sigma_{\lambda} \else $\sigma_{\lambda}$ \fi} 
\def \siglamp {\ifmmode                \sigma_{\lambda^{\prime}} \else
$\sigma_{\lambda^{\prime}}$\fi}

\def \Rd {\ifmmode R_{\rm d} \else $R_{\rm d}$ \fi} 
\def \Rs {\ifmmode R_{\rm s} \else $R_{\rm s}$ \fi}  
\def \Rd {\ifmmode R_{\rm d} \else $R_{\rm d}$ \fi}  
\def \Rcool  {\ifmmode R_{\rm  cool}  \else $R_{\rm cool}$ \fi} 
\def \RIII {\ifmmode  3.2\Rs \else $3.2\Rs$ \fi} 
\def \RII {\ifmmode 2.2\Rs \else $2.2\Rs$  \fi} 
\def \Reff {\ifmmode R_{\rm eff} \else $R_{\rm  eff}$ \fi} 
\def  \rb {\ifmmode r_{\rm b}  \else $r_{\rm b}$ \fi}

\def  \Sigmacrit   {\ifmmode  \Sigma_{\rm  crit}   
\else  $\Sigma_{\rm crit}$\fi} 
\def \Sig0 {\ifmmode \Sigma_{0} \else $\Sigma_{0}$ \fi}

\def \muI {\ifmmode \mu_{0,I} \else $\mu_{0,I}$ \fi}

\def \mgal {\ifmmode m_{\rm gal} \else $m_{\rm gal}$ \fi} 
\def \md {\ifmmode m_{\rm d} \else $m_{\rm d}$ \fi} 
\def \ms {\ifmmode m_{\rm   s}   \else   $m_{\rm   s}$   \fi}   
\def   \mdbar   {\ifmmode {\overline{m}}_{\rm d} \else
  ${\overline{m}}_{\rm d}$ \fi} 
\def \msbar {\ifmmode  \bar{m}_{\rm  s}  \else  $\bar{m}_{\rm s}$
  \fi}  
\def  \Md {\ifmmode M_{\rm d}  \else $M_{\rm d}$ \fi} 
\def  \Ms {\ifmmode M_{\rm s} \else $M_{\rm  s}$ \fi} 
\def \Mb {\ifmmode  M_{\rm b} \else $M_{\rm b}$ \fi} 
\def \Mstar {\ifmmode  M_{\rm star} \else $M_{\rm star}$ \fi}
\def \Mdisc {\ifmmode M_{\rm disc} \else $M_{\rm disc}$ \fi}

\def \Jd {\ifmmode J_{\rm d} \else $J_{\rm d}$ \fi} 
\def \Jb {\ifmmode J_{\rm b} \else $J_{\rm b}$ \fi}  
\def \fb {\ifmmode  f_{\rm b} \else $f_{\rm b}$ \fi}

\def  \jd  {\ifmmode j_{\rm  d}  \else  $j_{\rm  d}$ \fi}  
\def  \jdmd {\ifmmode \frac{j_{\rm  d}}{m_{\rm d}} \else
  $\frac{j_{\rm d}}{m_{\rm d}}$ \fi} 
\def \fj {\ifmmode f_{\rm j} \else $f_{\rm j}$ \fi} 
\def \ft {\ifmmode f_{\rm t}  \else $f_{\rm t}$ \fi} 
\def  \fM {\ifmmode f_{\rm M} \else $f_{\rm M}$ \fi}

\def  \Vd {\ifmmode  V_{\rm  d}  \else $V_{\rm  d}$  \fi} 
\def  \Vcool {\ifmmode V_{\rm cool} \else $V_{\rm cool}$ \fi} 
\def \Vcirc {\ifmmode V_{\rm circ}  \else $V_{\rm circ}$  \fi} 
\def \VIII  {\ifmmode V_{3.2} \else $V_{3.2}$ \fi} 
\def  \VII {\ifmmode V_{2.2} \else $V_{2.2}$ \fi}
\def \Vobs {\ifmmode V_{\rm obs}  \else $V_{\rm obs}$ \fi} 
\def \Vdisc {\ifmmode V_{\rm disc} \else  $V_{\rm disc}$ \fi} 
\def \Vmax {\ifmmode V_{\rm  max} \else  $V_{\rm max}$  \fi} 
\def  \Vmaxobs{\ifmmode V_{\rm max}^{\rm obs}\else  $V_{\rm max}^{\rm
    obs}$\fi}  
\def \Vtot {\ifmmode V_{\rm tot} \else $V_{\rm tot}$  \fi} 
\def \Vrot {\ifmmode V_{\rm rot} \else  $V_{\rm rot}$  \fi} 
\def  \Vflat {\ifmmode  V_{\rm  flat} \else $V_{\rm flat}$ \fi}

\def \Ups {\ifmmode \Upsilon  \else $\Upsilon$ \fi} 
\def \YB {\ifmmode \Upsilon_B \else $\Upsilon_B$ \fi} 
\def \YI {\ifmmode  \Upsilon_I  \else $\Upsilon_I$ \fi} 
\def \DeltaIMF {\ifmmode \Delta_{\rm IMF} \else $\Delta_{\rm IMF}$ \fi}

\def\LCDM{$\Lambda$CDM }

\def\c200{$c_{200}$}

\begin{document}
\submitted{The Astrophysical Journal, submitted}
\vspace{1mm}
\slugcomment{{\em The Astrophysical Journal, submitted}}

\shortauthors{MACCI\`O ET AL.}
\twocolumn[
\lefthead{Halo expansion in hydro simulations}
\righthead{Macci\`o et al.}

\title{Halo expansion in cosmological hydro simulations: towards a baryonic solution of the cusp/core 
problem in massive spirals}

\author{A. V. Macci\`o\altaffilmark{1}, G. Stinson\altaffilmark{1,2}, C.B. Brook\altaffilmark{2},
J. Wadsley\altaffilmark{3}, H.M.P. Couchman\altaffilmark{3}, S. Shen\altaffilmark{4}, \\
B.K. Gibson\altaffilmark{2,5}, T. Quinn\altaffilmark{6}
}

\begin{abstract}

A clear prediction of the Cold Dark Matter model is the
existence of cuspy dark matter halo density profiles on all mass scales.
This is not in agreement with the observed rotation
curves of spiral galaxies,  challenging on small scales the otherwise successful CDM paradigm.
In this work we employ high resolution cosmological hydro-dynamical simulations
to study the effects of dissipative processes on the inner distribution
of dark matter in Milky-Way like objects ($M\approx 10^{12} \Msun$).
Our simulations include supernova feedback, and the effects
of the radiation pressure of massive stars \emph{before} they explode as supernovae.
The increased stellar feedback results in the {\it expansion} of the dark matter halo instead of contraction with respect to N-body simulations. 
Baryons are able to erase the dark matter cuspy distribution creating a flat, cored,
dark matter density profile in the central several kpc of a massive Milky-Way like halo. 
The profile is well fit by a Burkert profile, with fitting parameters consistent with the observations.
In addition, we obtain flat rotation curves as well as extended, exponential stellar disk profiles.
While the stellar disk we obtain is still partially too thick to resemble the MW thin disk, this pilot
study shows that there is enough energy available in the baryonic component to 
alter the dark matter distribution even in massive disc galaxies, providing a possible solution to the long standing problem of cusps vs. cores.

\end{abstract}

]

\altaffiltext{1}{Max-Planck-Institut f\"ur Astronomie, K\"onigstuhl 17, 69117
  Heidelberg, Germany; maccio@mpia.de, stinson@mpia.de}
\altaffiltext{2}{University of Central Lancashire, Jeremiah Horrocks Institute for Astrophyics \& Supercomputing, Preston PR1 2HE}
\altaffiltext{3}{Department of Physics and Astronomy, McMaster University, Hamilton, Ontario, L8S 4M1, Canada}
\altaffiltext{4}{Department of Astronomy and Astrophysics, University of California, Santa Cruz, 1156 High Street, Santa Cruz, CA 95064}
\altaffiltext{5}{Department of Astronomy \& Physics, Saint Mary's University, Halifax, Nova Scotia, B3H 3C3, Canada}
\altaffiltext{6}{Astronomy Department, University of Washington, Box 351580, Seattle, WA, 98195-1580}

\section{Introduction}
\label{sec:intro}

The theory of  cold dark matter (CDM) provides  a successful framework
for understanding  structure formation in the universe  (e.g. Spergel \etal 2003, Komatsu \etal 2011). In this
paradigm, dark matter first collapses into small haloes, which merge to
form progressively  larger haloes.  Galaxies are thought to
form out of gas which cools and collapses to the centres of these dark
matter  haloes  (White  \&  Rees  1978).
Properties of dark matter haloes have been extensively studied via collisionless $N$-body simulations  
of the growth of primordial fluctuations into gravitationally bound structures.
Numerical simulations have facilitated
detailed predictions  for a wide range of properties of dark matter haloes at all 
epochs (e.g. Macci\`o \etal 2008, Prada \etal 2011).

Dissipationless cosmological simulations have also raised problems 
for the CDM scenario on small scales, one of which  is the central slope
of the dark matter density profile of virialized objects.
$N$-body simulations predict a central concentration, with a logarithmic slope of $\approx -1$ 
(Navarro, Frenk \& White 1997, Diemand \etal 2005, Springel \etal 2008). Such a `cuspy' matter distribution  is not supported by  observations 
of the rotation curves of spiral galaxies, which have  revealed that the dark halos encompassing 
 disc galaxies have a constant density core (e.g. Salucci \& Burkert 2000, Oh \etal 2008, 
Spano \etal 2008, Kuzio de Naray \etal 2009). Donato \etal (2009) 
have recently analyzed the rotation curves of a large sample of galaxies ranging over
all Hubble types and with luminosity as high as $M_B\approx -22$. Their analysis shows that the central surface density, is 
nearly constant and independent of galaxy luminosity.
 This issue presents a major  challenge for the otherwise successful CDM cosmological model.

By construction, dissipationless simulations do not include baryons. While on large scales, 
the effect of gas and stars can be neglected, this is not true on small scales, where
baryons can be gravitationally dominant.
For this purpose, cosmological hydro-dynamical simulations have been extensively used 
to directly address the question of galaxy properties in the CDM scenario
(e.g. Navarro \& Steinmetz 2000, Abadi \etal 2003, Brook \etal 2004, Robertson \etal 2004, Okamoto \etal 2005, 
Macci\`o \etal 2006, Governato \etal 2007, S{\'a}nchez-Bl{\'a}zquez \etal 2009,
Scannapieco \etal 2009, Stinson \etal 2010, Piontek \& Steinmetz 2011, Agertz \etal 2011).

The response of dark matter to baryonic infall (and star formation) is still highly debated.
During galaxy formation, as cosmic gas cools and condenses towards the halo centre 
and forms stars, dark matter particles are pulled inward and increase their central density. This process  is dubbed
`adiabatic contraction' (e.g. Blumenthal \etal 1986, Gnedin \etal 2004).
Halo contraction is present  in the vast majority of cosmological hydro-dynamical simulations 
(see Gnedin \etal 2011 and references therein),
yet these simulated galaxies fail to reproduce observed rotation curves due to their too 
centrally concentrated stellar and dark matter profiles. In order to reproduce
observational data, like the rotation velocity-luminosity and size-luminosity relations,
models with NO adiabatic contraction  (Gnedin \etal 2007, Courteau \etal 2007) 
or even with expansion (Dutton \etal 2007, 2011) are required  (but also see Trujillo \etal 2010).

To resolve these discrepencies, theoretical arguments and  simulations have proposed baryonic processes that can result in producing an  expansion of the dark matter halo.
Gas bulk motions, possibly supernova-induced in regions of high star formation activity,
and the subsequent energy loss of gas clouds due to dynamical friction can 
transfer energy to the central dark-matter component (Navarro, Eke \& Steinmetz 1996, Mo and Mao 2004, El-Zant \etal 2001,Ogiya 
\& Mori 2011).

Mashchenko \etal (2006, 2008) have pointed out that there is another (possibly more relevant) effect, namely 
the gas bulk motion can induce substantial gravitational potential fluctuations 
and a subsequent reduction in the central dark matter density.
Recently Governato \etal (2010) have presented high resolution cosmological simulation of a dwarf galaxy that was able to create a cored dark matter profile
at $z=0$ within the CDM scenario, and reproduce several properties of observed dwarf galaxies.   
Resolution of an inhomogeneous interstellar medium, and   strong supernova driven outflows, which inhibits the formation of bulges, 
resulted in the decrease of the dark-matter density to less than half of what it would otherwise be within the central kiloparsec of 
these low mass objects ($M_{dm} \approx 10^{10} \Msun$).
Further analysis of this simulation, Pontzen \& Governato (2011)  showed that 
the flattening was the result of relatively small starbursts in the centre of the protogalaxy, which contribute over many cycles to a gradual transfer of energy from the baryons to the dark matter.
This mechanism is closely related to matter outflows, but does not require violent, sudden mass loss. 
In a recent paper Brook \etal (2011) showed that large fractions of the gas that is expelled from the central regions of galaxies returns via a large scale 
galactic fountain to form stars at later times: this greatly increases the occurrence of outflows from the inner regions for a galaxy of given stellar mass, and thus any flattening mechanism that relates to outflows will be significantly enhanced.

Dekel \& Silk (1986) showed that supernovae can eject matter from halos up to 100 km s$^{-1}$, but it has yet to be seen what impact this might have on dark matter profiles, 
nor how the addition of radiation pressure feedback might change things.  
So, while observations show evidence for flattened dark matter density profiles up to L* galaxies, the question remains whether there is enough energy input from baryons in more massive objects in 
order for these processes to be effective in altering the dark matter density profile of spiral galaxies with a dark matter mass of the order of $10^{11}-10^{12}$.

In this letter we present a  high resolution cosmological hydrodynamic simulation of a massive spiral galaxy that includes cooling due to hydrogen and heavier "metal" elements (Shen \etal 2010), UV background radiation (Haardt \& Madau from \textsc{cloudy}), a simple commonly used star formation prescription, adiabatic supernova feedback 
(Stinson \etal 2006), along with feedback from the early radiation produced by massive stars (Stinson \etal in prep, see Brook \etal 2011).
Our results show that reasonable baryonic feedback is able to create a density 
core in the dark matter distribution even for massive spiral galaxies approaching the mass of our own Milky Way.

This letter is organized as follows in Section \ref{sec:sims} we describe our simulations, in Section \ref{sec:res}
we present our results on the dark matter radial density profile, finally, Section \ref{sec:concl} is devoted to conclusions and 
discussion.

\section{Simulations}
\label{sec:sims}

In  this paper we use g5664, a cosmological zoom simulation drawn
from the McMaster Unbiased Galaxy Simulations (MUGS).  See Stinson \etal 2010
for a complete description of the creation of the initial conditions.  g5664
has a total mass of $7\times10^{11}$ M$_\odot$, a spin parameter of 0.024, and
a last major merger at $z=3.4$.  Inside $r_{vir}$ at $z=0$, there are $4\times 10^5$ 
dark matter particles with a mass of  $1.1\times10^{6}$ M$_\odot$, a similar number gas particles with mass $2.1\times10^{5}$ M$_\odot$ 
and between $3.5\times10^5$ and $10^6$ stars with mass $5.5\times10^{5}$ M$_\odot$ depending on the star formation
and feedback recipes used.
Using the physics employed in the original MUGS 
simulations, g5664 formed an exponential disk with an exponential bulge 
with a central surface brightness of $\mu_i=18$, had a total face-on magnitude 
of M$_r$=-21.7 and $g-r$ colour of 0.52.  In many ways, g5664 is a similar initial condition 
to the Guedes \etal (2011) \emph{eris} simulation, except 8 times lower resolution.
We will refer to this simulation as the low feedback run (LFR).

The new version of g5664  was also evolved using the smoothed 
particle hydrodynamics (SPH) code \textsc{gasoline} (Wadsley \etal 2004).
For this high feedback run (HFR), three changes were made which resulted in a stronger implementation of stellar feedback:
i) The Kroupa \etal (1993) IMF used in MUGS was changed to the more commonly used Chabrier (2003) IMF which creates more massive stars hence more energy
per stellar mass created.  
ii) We use a star formation density threshold of $n_{th}$ 9.3 cm$^{-3}$ and an efficiency of $c_\star=0.1$.
iii) The new runs assume the energy input from supernovae
is $10^{51}$ ergs instead of the $4\times10^{50}$ ergs used in MUGS. 
iv) the new run includes radiation released by the massive young stars before they
explode as supernovae.  
All of these changes are based on a parameter search that will be reported in Stinson \etal (in prep) 
It will be shown that each of these changes is necessary to produce more realistic galaxies as described in \S \ref{sec:res}.
For details on the implementation of the radiative 
feedback, see Brook \etal (2011), though the work presented here deposits 
17.5\% of the massive star luminosity as thermal energy instead of the 10\% 
reported there.  
Again, this is motivated by the production of a more realistic galaxy.
We note here that the energy deposition from supernovae and radiation pressure couple inefficiently 
to the gas. Both types of feedback are deposited into the high density gas found in the disk where the cooling time is shorter than the dynamical timescales resolved in the simulation. 

We emphasise that our feedback scheme relies on subgrid physics.  It is not 
possible to fully resolve the sites where star formation happens and energy is fed back into the ISM. Our simple description of the radiation pressure mechanism is a relatively  crude initial representation.  Finally, we also ran a dark matter only version of g5664  (N-body run), with the same particle resolution.

\section{Results}
\label{sec:res}

In this Letter we  focus on the effects of  feedback on 
the dark matter distribution in our simulated galaxy. A more comprehensive 
study of the properties of simulated galaxies using such a strong feedback implementation will be presented in a  forthcoming series of papers.  In brief, the simulation presented here does a much better job reproducing
the observed properties of galaxies including a flatter rotation curve, an exponential  surface
brightness profile and a stellar mass, $6\times10^9$ M$_\odot$, in better agreement with 
what halo abundance matching predicts a $7\times10^{11}$ M$_\odot$ should contain
(Moster \etal 2010).  This latter point is important, as it indicates that the large scale outflows inherent in this study may be {\it necessary}.

\begin{figure}[t]
\centerline{\epsfxsize=3.2in \epsffile{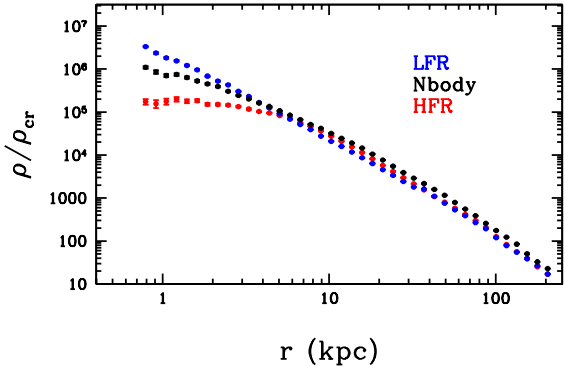}}
\caption{ Density profile for only the dark matter in the three different realizations of our galaxy.
The \emph{blue} line shows the low feedback run (LFR), the \emph{black} line shows the dark matter only 
run (N-body) and the \emph{red} line shows the higher
feedback case. The blue curve shows evidence for adiabatic contraction, the black one presents the 
usual NFW profiles, while the red one shows a clearly a cored profile, in agreement with observations.}
\label{fig:profall}
\end{figure}

Figure \ref{fig:profall} shows the dark matter density profile in our three simulations. The pure dark matter run (N-body, black line) is well fit by an NFW profile with concentration parameter of $c=6$, in agreement with cosmological expectations (Mu{\~n}oz-Cuartas \etal 2011).
The profile of the low feedback run  \emph{(LFR, dotted-blue)}   shows evidence of significant adiabatic contraction, with DM pulled towards the inner regions by the centrally concentrated baryons. The inner profile is fit with a single power
law ($\rho \propto r^{-\alpha}$), with $\alpha=2$.
As reported in Stinson \etal (2010), this dark matter peak is accompanied by a high concentration of baryonic material at the centre 
of the galaxy, represented by the a centrally peaked rotation curve and high bulge-to- total ratio.
None of these features agree with observations, which do not support the adiabatic contraction scenario at these mass scales.
The lowest curve is our HFR, which uses a Chabrier IMF and radiation pressure feedback.
The dark matter density profile follows the pure dark matter run in to $r\approx 5 kpc$, but then it notably flattens to clearly reveal the presence of a core in the inner region.

\begin{figure}[t]
\centerline{\epsfxsize=3.2in \epsffile{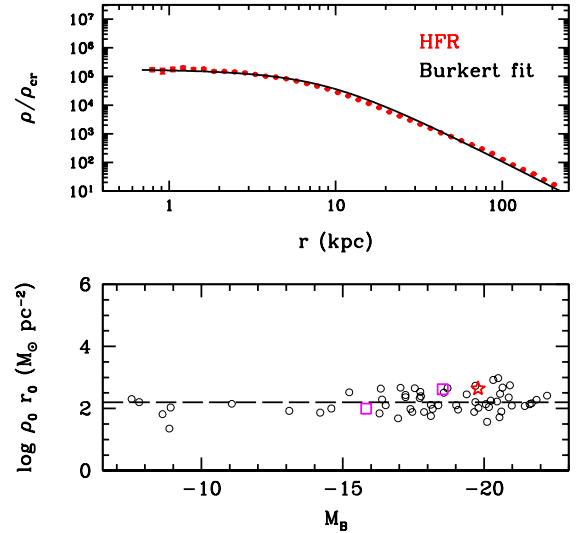}}
\caption{\emph{Upper panel}: density profile of the DM component in g5664 HFR, 
and  fitting Burkert profile with a core size of $r_0=9.11$ kpc. \emph{Lower panel}: The relation between luminosity and dark matter halo surface density. 
Open symbols represent observational results, while our simulated galaxy is represented by the 
red star. The dashed line is the fit to this relation, suggested by Donato \etal (2009).  The magenta squares are lower mass simulations which have similar high feedback prescriptions as the HFR 
(Brook \etal 2011b in prep).}
 \label{fig:profb2}
\end{figure}

The DM density profile of the HFR can be fit with a Burkert profile (Burkert 1995):
\begin{equation}
\rho(r) = { {\rho_0r^3} \over {(r+r_0) \left(r^2+r^2_0 \right)} }.
\end{equation}
This profile, when combined with appropriate baryonic gaseous and stellar components, 
is found to reproduce very well the oserved kinematics of disc systems (e.g. Salucci \& Burkert 2000; Gentile \etal 2007).
The two free parameters $(\rho_0;r_0)$ can be determined through a $\chi^2$ minimization fitting procedure:
In our case this led to $\rho_0/\rho_{cr} = 1.565 \times 10^5$ and $r_0 =9.11$ kpc.
The simulated DM profile with its Burkert fit are shown in the upper panel of 
Figure \ref{fig:profb2}. 

 Results  of Donato \etal (2009), showing the central surface density $\mu_0$, defined as
the product of the halo core radius and central density ($\mu_0 \equiv r_0 \rho_0$) of galactic dark matter haloes, are shown (open circles) in the lower panel of Figure \ref{fig:profb2}.
Our simulated galaxy is over-plotted as a red star. Not only can the simulation be fit with a Burkert profile, but the cored profile of g5664 HFR agrees with observed density profiles. 
The magenta squares are lower mass simulations which have similar high feedback prescriptions as the HFR, with slight calibration changes as these simulations  have 8 times better resolution. 
Detailed properties of these simulations will be presented in a forthcoming paper (Brook \etal 2011b in prep).

\subsection{When and how is the density profile flattened?}
\begin{figure}[t]
\centerline{\epsfxsize=3.2in \epsffile{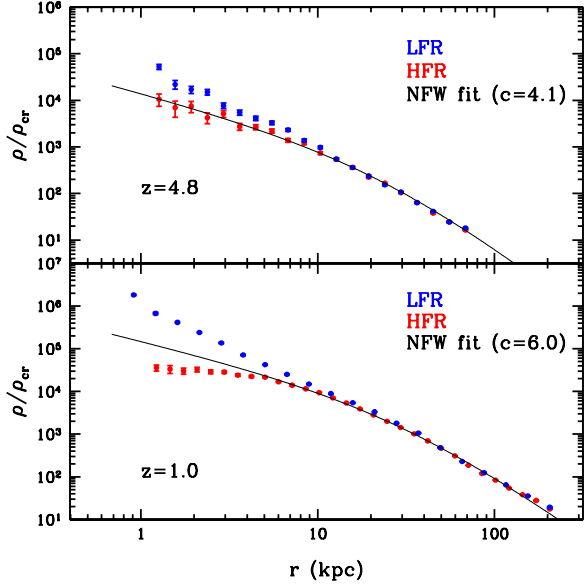}}
\caption{ Redshift evolution of the dark matter density profile. In both panels
the NFW fit has been obtained by fixing the concentration $c_{vir}=4.1$ and $c_{vir}= 6.0$ at $z=4.8$ and 1.0 respectively, 
according to results of Mu{\~n}oz-Cuartas \etal 2011 based on N-body cosmological simulations. }
\label{fig:profdmz}
\end{figure}

Figure \ref{fig:profdmz}  shows the dark matter density profile for the hydrodynamical simulations 
(low and high feedback) at 
$z=4.8$ and $z=1$ (upper and lower panel respectively). The two runs show markedly different behaviours:
High feedback results in low star formation rates in low mass progenitors, as it prevents significant gas cooling to the very central regions of the dark matter halos. The
dark matter profile remains unperturbed from pure N-body simulations (black solid line). In the low feedback case, gas cools rapidly to the central regions at high $z$, and the  dark matter adiabatically contracts. At $z=1$ (lower panel) the energy transfer from gas to dark matter in the HFR has already considerably
flattened the density profile of this latter component, that now clearly deviates from 
$N$-body based expectations. The profile of the MUGS run (LFR) is still contracted and has reached a
logarithmic slope of $\alpha = 2$.

\begin{figure}[t]
\centerline{\epsfxsize=3.2in \epsffile{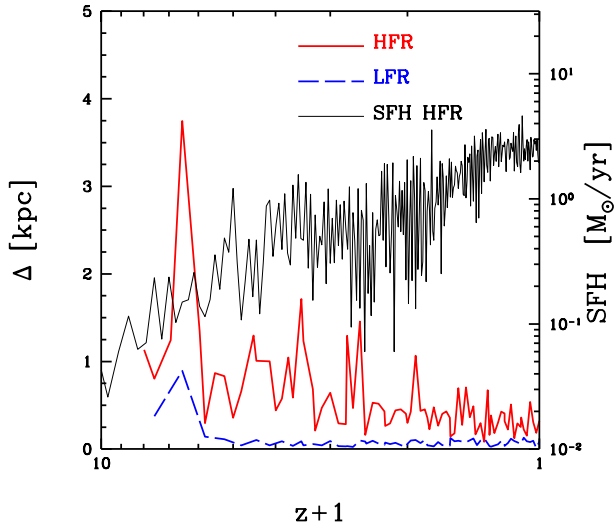}}
\caption{Evolution of the distance between the position of the dark matter and gas 
potential minima. The solid (red) line and the dashed (blue) line represent the high and low feedback case respectively.
The thin black line shows the Star Formation History for the high feedback run.}
\label{fig:minpotsfr}
\end{figure}

The creation of a core in the dark matter distribution has previously been  attributed to rapid variations on the potential due to the bulk motion of gas clouds (Mashchenko \etal 2008, Pontzen \& Governato 2011). In Figure \ref{fig:minpotsfr} we quantify this variation by plotting the distance, $\Delta$, between the position of the most bound dark matter ($\vec x_{DM}$) and gas ($\vec x_{gas}$) particles.
In the HFR (red line) the potential is rapidly changing potential is reflected in the oscillations of , $\Delta$  with time,  with the amplitude of the oscillations of the order of the size of the dark matter core ($\approx$ kpc).
In the LFR, the roughly constant and small value of  $\Delta$ indicates a more stable potential.  
This indicates that the  changing potential is responsible for generating  dark matter cores in our HFR.
We note that while bulk gas motions are a natural result of star formation and feedback,
it is harder to conceive how such a mechanism would work with AGN feedback.

\section{Summary and Conclusions}
\label{sec:concl}

Several mechanisms have been proposed to flatten dark matter profiles, in order to reconcile the tensions between the observed cored profiles and the cusps of CDM predictions. Fully cosmological simulations have been able to show that such processes can occur in low mass (dwarf galaxy) systems (Governato 2010). Yet cosmological simulations of more massive disc galaxies have invariably resulted in adiabatic contraction (Gnedin \etal 2011), contradicting observed disc galaxies (Oh \etal 2008, Donato \etal 2009). The centrally peaked rotation curves, and high stellar mass fractions of these simulated galaxies  also fail to reproduce observations.

In this work we have, for the first time, explicitly considered the feedback 
from radiation pressure due to massive stars in a cosmological hydro-dynamical simulation of galaxy formation.
We compared the dark matter density profile of this new simulation with a twin run that only 
considered relatively weak feedback from Supernovae.
We have explicitly shown that stronger stellar feedback can reverse the effect of adiabatic contraction, 
and {\it expand} dark matter halos  massive enough to host L$_*$ galaxies .  The cored profiles in our simulated galaxy have a core radius and central density that agrees with observations (see fig. \ref{fig:profb2}). Simultaneously, the high feedback  simulations have rotation curves and stellar masses that are also better matches to observed disc galaxies than their low feedback counterparts. 

The flattening of the profile is due to the fluctuation of the global potential, both in its depth 
(Pontzen \& Governato 2011) and its position (fig. \ref{fig:minpotsfr}).  The flattened profile arises at intermediate redshifts, when 
strong star formation and subsequent energy injection from feedback in shallower potential wells has the strongest effect. At high redshift ($z \approx 3$)
the dark matter density profile is still in agreement with NFW-like predictions (Mu{\~n}oz-Cuartas \etal 2011).

This pilot study shows  that, with reasonable baryonic feedback, there is enough energy input in the central region of the galaxy to induce rapid change 
in the potential and, eventually, induce a  dark matter halo expansion, helping in reconciling observations with CDM predictions. We emphasise that, 
while the simulation presented here does a good job in reproducing the observed dark matter profiles, it does not addresses all aspects of the galaxy formation.

We are working on  expanding our current work to explore the effects of our feedback on a range of 
properties of galaxies in a wide range of masses (e.g. the two objects already shown in fig. \ref{fig:profb2}). 
The case of  Low Surface Brightness galaxies, in particular, will be interesting since they also present cored density profiles yet have very low baryon fractions.
Further, higher resolution simulations will be required to confirm the impact of stellar feedback on the formation of density profiles 
in galaxies of varying masses.  However, we hope this work represents a first step down a new path to creating more realistic galaxies.

\acknowledgements

We thank Aaron Dutton for helpful comments on an early version of this manuscript.
AVM also thanks P. Salucci and F. Donato for sending an electronic version 
of their data points and Brent Groves for useful conversations.
Numerical simulations were performed on the THEO cluster of  the
Max-Planck-Institut f\"ur Astronomie at the Rechenzentrum in Garching, 
on the Universe cluster run by COSMOS in Cambridge, 
on the SHARCNET clusters in Canada, and on the University of Central Lancashire High Performance Computing Facility.
AVM acknowledges funding by Sonderforschungsbereich SFB 881 ``The Milky Way System'' 
(subproject A1) of the German Research Foundation (DFG).
BKG acknowledges the support of the UK's Science \& Technology Facilities Council (
ST/F002432/1 \& ST/H00260X/1).


\end{document}